\documentclass[twocolumn,superscriptaddress,aps,prl,showkeys]{revtex4-1}
\usepackage[utf8]{inputenc}
\usepackage{color}
\usepackage[dvipsnames]{xcolor}
\usepackage{amsmath}
\usepackage{amssymb}
\usepackage{amsbsy}
\usepackage{bbold}
\usepackage{graphicx}
\usepackage{xfrac}
\usepackage{placeins}
\usepackage[]{todonotes}
\usepackage{lineno}
\usepackage[unicode=true,bookmarks=false,
 breaklinks=false,pdfborder={0 0 1},backref=false,colorlinks=true]{hyperref}
\hypersetup{
 linkcolor=blue,citecolor=blue,urlcolor=blue,linkcolor=blue,citecolor=blue,urlcolor=blue}


\begin{document}
% \linenumbers

\title{Multiplexed chiral and helical acoustic anomaly bulk states in an inverse-designed metamaterial}

\author{Xueyun Wen}
\affiliation{Institute of Acoustics, School of Physics Science and Engineering, Tongji University, Shanghai 200092, China}
\author{Zhihao Lan}
\affiliation{College of Physical Sciences and Engineering, Mohammed VI Polytechnic University, Ben Guerir, 43150, Morocco}

\author{Zhongming Gu}

\affiliation{Institute of Acoustics, School of Physics Science and Engineering, Tongji University, Shanghai 200092, China}

\author{Yafeng Chen}
\email{yachen@tongji.edu.cn}
\affiliation{Institute of Acoustics, School of Physics Science and Engineering, Tongji University, Shanghai 200092, China}

\author{Jie Zhu}
\email{jiezhu@tongji.edu.cn}
\affiliation{Institute of Acoustics, School of Physics Science and Engineering, Tongji University, Shanghai 200092, China}
\affiliation{Shanghai Research Institute for Intelligent Autonomous Systems, Tongji University, Shanghai 201210, China}

\begin{abstract}
Valley and pseudospin are fundamental degrees of freedom (DOFs) in topological acoustics.
However, they have so far been realized mostly in separate systems. Here, we achieve the multiplexing of valley and pseudospin DOFs in one acoustic system. An acoustic metamaterial simultaneously hosting single and double Dirac cones is inversely designed by the developed topology optimization method.
Thereafter, under a hard-boundary condition, valley-locked chiral states and pseudospin-locked helical states emerge within two separate frequency windows and can be selected by tuning the operating frequency.
Furthermore, their applications for multifrequency acoustic energy enhancement are experimentally demonstrated. Our work provides a strategy to modulate acoustic waves carrying distinct topological DOFs within a single integrated platform, facilitating the development of multiplexed acoustic devices with robustness. 
\end{abstract}

% \keywords{anomaly bulk state; quantum valley Hall effect; quantum spin Hall effect.}

\maketitle

\section{Introduction}
The discovery of topological insulators (TIs), characterized by topologically protected states with strong robustness and ultra-low energy dissipation, has opened a new chapter in condensed-matter physics, holding tremendous promise for the development of high-efficiency spintronic devices and quantum information processors \cite{1Hasan,2Qi,3Pesin,4Breunig,5HeSpintronics}.
As the efficient and robust manipulation of acoustic waves has potential applications for developing novel devices, the concept of TIs has also been transplanted into acoustic systems \cite{6Mousavi,7Khanikaev,8Susstrunk,9XiaoGeo,10XiaoWeyl,11YangAcoustics,12Huber,13ZhangSound,14MaReview,15XueReview}.
However, due to the lack of intrinsic spin-1/2 behavior of electrons, it needs to synthesize artificial degrees of freedom (DOF) to mimic quantum effects of TIs.
While various artificial DOFs have been engineered in acoustic systems to construct acoustic TIs, the valley and pseudospin DOFs used to mimic quantum valley Hall effects (QVHEs) and quantum spin Hall effects (QSHEs) are the most prominent \cite{16HeATI,17ZhangMultipoles,18DengSpinChern,19LuValley,20LuBilayer,21WangSAW,22He3DTopo,23SunQSH,24He3DTI}.
By breaking spatial inversion symmetry to lift the degeneracy of the single Dirac cones at the $K$/$K^\prime$ points and open a bandgap, the valley DOFs can be introduced and the interface between two domains with opposite valley Chern number can host valley-locked chiral edge states \cite{19LuValley,20LuBilayer,25GaoValleyChern,26YeVortex}.
On the other hand, pseudospin DOFs can be constructed by taking linear combinations of the degenerate dipole and quadrupole modes that form a double Dirac cone at the Brillouin zone center. By lifting the degeneracy of the double Dirac cone to induce a band inversion between the dipole and quadrupole modes, trivial and nontrivial unit cells (UCs) with opposite spin Chern number can be designed and pseudospin-locked helical edge states can emerge at the interface between them \cite{16HeATI,18DengSpinChern}.
Besides, by sandwiching a domain of UCs with single (double) Dirac cone between two topologically distinct domains, large-area valley-locked (pseudospin-locked) waveguide states are constructed, for which the mode localization area can be modulated by tuning the number of Dirac cone layers \cite{21WangSAW,27LanLargeArea}.
In addition to constructing valley- and pseudospin-locked interface states within the bandgap by lifting Dirac cones, one can also achieve valley-locked chiral (pseudospin-locked helical) anomaly bulk states by engineering the boundaries of metamaterials composed of UCs with a single (double) Dirac cone \cite{28LiCABS,29WangCABS,30ZhangFreeBoundary}.

The introduction of valley and pseudospin DOFs into acoustic systems has endowed the scalar acoustic waves with quantum-like properties, such as valley/ pseudospin-locked unidirectional transmissions that are immune to defects, overturning traditional views on the manipulation of acoustic waves \cite{31DengZoneFolding,32MeiPTR}.
Although several works have attempted to combine these two prominent DOFs, the valley and the pseudospin have either originated from a common parent degeneracy, so that the two indices are not independent \cite{22He3DTopo}, or have been produced by imposing distinct structural perturbations on a shared platform, so that two structurally inequivalent UCs are still required \cite{WangPRB2019}.
Moreover, all these previous realizations were based on topological edge states, and to our knowledge, the integration of both valley and pseudospin acoustic anomaly bulk states into a single acoustic platform has not been reported thus far.
Given the fundamental importance of valley and pseudospin DOFs, it is highly desirable to integrate them into a single acoustic platform, thereby doubling the capacity for acoustic information processing. 
However, this remains challenging: the two degeneracies must not only coexist, but also occur at prescribed frequencies with spectrally isolated neighborhoods. These coupled spectral constraints are difficult to satisfy simultaneously by tuning only a small set of correlated geometric parameters.

\begin{figure}[h!]
\centering
\includegraphics[width=\columnwidth]{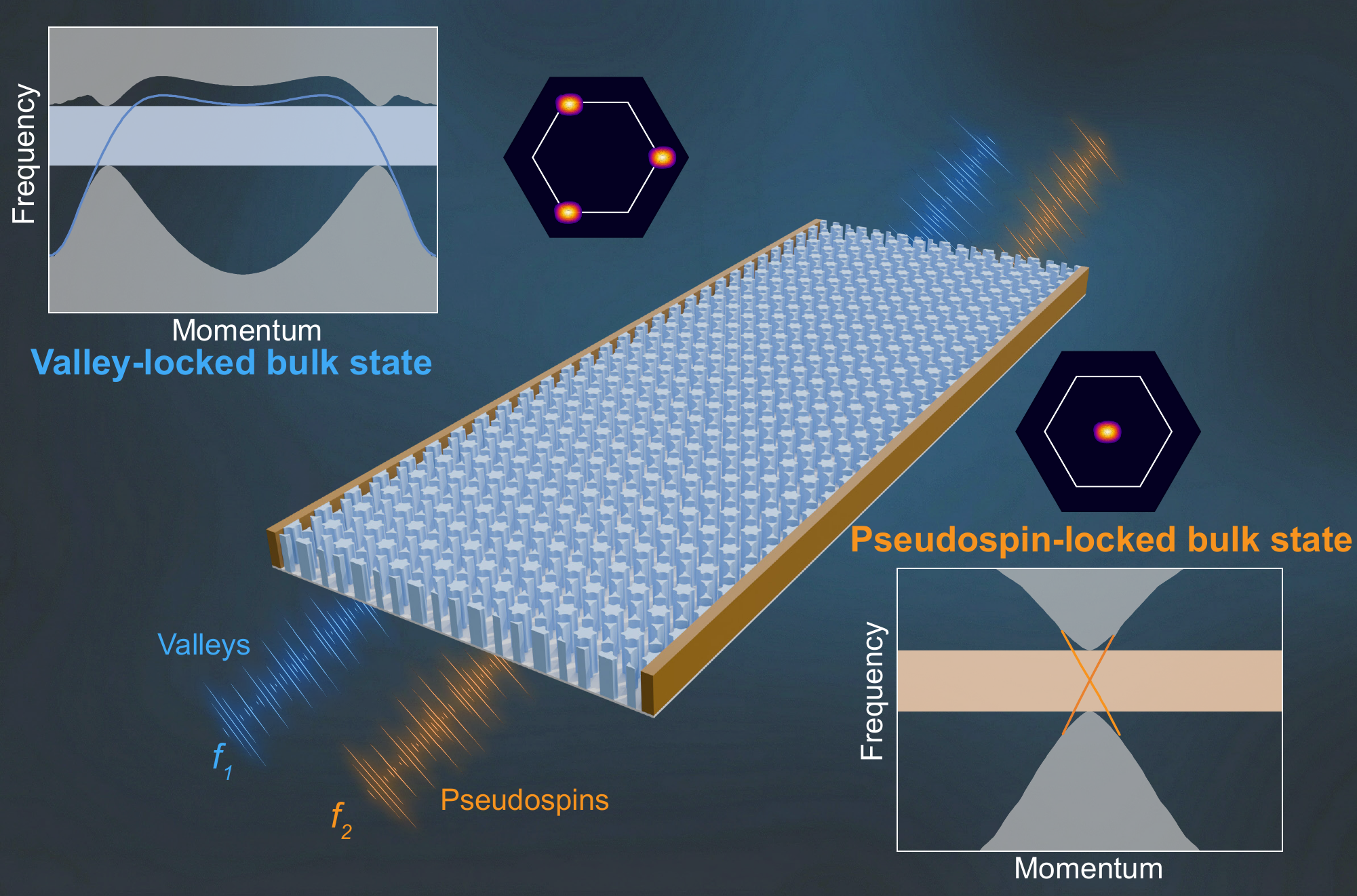} 
	\caption{The schematic of the metamaterial hosting multiplexed chiral and helical acoustic anomaly bulk states with pseudospins and valleys.}
	 \label{fig:fig1}
\end{figure}

In this work, we achieve the integration of valley and pseudospin DOFs within a single acoustic metamaterial.
An acoustic crystal hosting both single and double Dirac cones is inversely designed via the developed topology optimization method.
Through boundary engineering, we achieve multiplexed chiral and helical acoustic anomaly bulk states that carry both valleys and pseudospins, which can be switched via tuning the frequency, as conceptually illustrated in Fig. \ref{fig:fig1}.
We experimentally verify the dispersion distributions in momentum space of the valley- and pseudospin-locked anomaly bulk states and confirm their robustness. Furthermore, by exploiting the uniform phase distributions of these anomaly bulk states, we experimentally demonstrate an energy enhancement effect for acoustic waves traversing a step waveguide. Our work provides a platform for studying different topological DOFs and offers the opportunity for enhancing acoustic information processing capabilities by multiplexing data across distinct DOFs.

\section{Results}
As valley and pseudospin DOFs respectively originate from the single Dirac cone and the double Dirac cone, to construct a system that accommodates both DOFs, we first design a $C_{6v}$-symmetric UC that simultaneously hosts a single Dirac cone at the $K$/$K^\prime$ point and a double Dirac cone at the $\Gamma$ point using the developed topology optimization method [see Appendix A for details], as shown in Fig. \ref{fig:fig2}(a), where the blue and gray regions represent the solid material and the air domain, respectively.
The lattice constant is $a = 5$ cm.

\begin{figure}[h!]
\centering
\includegraphics[width=\columnwidth]{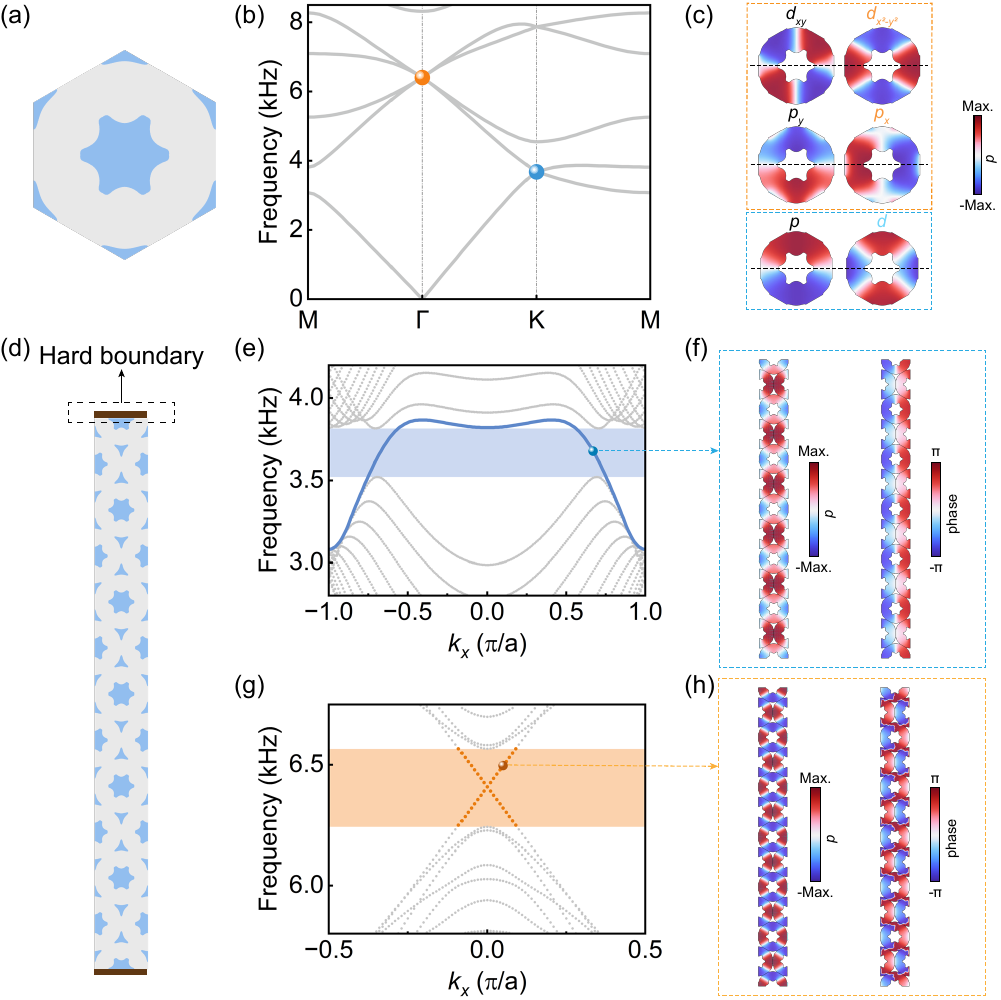} 
	\caption{(a) Schematic of the inverse-designed UC. (b) Band structure of the UC. The blue (orange) dot indicates the single (double) Dirac cone. (c) Corresponding acoustic pressure fields of the eigenmodes that form the single and double Dirac cones. (d) Schematic of the supercell with a hard boundary by truncating the outermost upper and bottom UCs along their horizontal center symmetry line. (e) Band dispersion of the supercell around the single Dirac cone. The shaded region denotes the working frequency window of the valley-locked anomaly bulk states. (f) The acoustic pressure (left) and phase (right) fields of the eigenmode marked by the blue ball in (e). (g) Band dispersion of the supercell around the double Dirac cone. The shaded region denotes the working frequency window of the pseudospin-locked anomaly bulk states. (h) The acoustic pressure (left) and phase (right) fields of the eigenmode marked by the orange ball in (g).}
	 \label{fig:fig2}
\end{figure}

The band structure of the UC, shown in Fig. \ref{fig:fig2}(b), confirms the appearance of a single Dirac cone (denoted by the blue ball) at the $K$ point and a double Dirac cone (denoted by the orange ball) at the $\Gamma$ point.
Figure \ref{fig:fig2}(c) shows the two degenerate eigenmodes ($p$/$d$, denoted by the blue dashed box) that form the single Dirac cone and the four degenerate eigenmodes ($d_{x^2-y^2}$/$d_{xy}$ and $p_x/p_y$, denoted by the orange dashed box) that form the double Dirac cone.
When the upper and bottom hard boundaries are imposed by truncating the supercell along the centerline of the outermost UCs [Fig. \ref{fig:fig2}(d)], the boundary condition enforces a vanishing normal velocity ($v_n = 0$).
Consequently, only bulk modes that are mirror-symmetric with respect to the $x$ axis ($d$/$p_x$/$d_{x^2-y^2}$), for which the normal velocity component vanishes on the mirror plane, are compatible with the hard-boundary condition, whereas mirror-antisymmetric modes ($p$/$p_y$/$d_{xy}$) with finite $v_n$ on the mirror plane are suppressed.
The corresponding parity-selection mechanism is discussed via a tight-binding model in Appendix B.
Figures \ref{fig:fig2}(e) and \ref{fig:fig2}(g) show the projected band dispersions of the supercell in Fig. \ref{fig:fig2}(d) around the single and double Dirac cones, respectively.
It can be observed that, after suppressing the band associated with the mirror-antisymmetric mode in the single Dirac cone, a QVHE-like (valley-locked) dispersive branch appears around the $K$ and $K^\prime$ points with the frequency window of 2.80--4.20 kHz [Fig. \ref{fig:fig2}(e)].
Meanwhile, with the two mirror-antisymmetric bands of the double Dirac cone suppressed, a pair of QSHE-like (pseudospin-locked) bands emerge around the $\Gamma$ point with the frequency range of 5.80--6.75 kHz [Fig. \ref{fig:fig2}(g)].
Figures \ref{fig:fig2}(f) and \ref{fig:fig2}(h) present the representative eigenmodes of the valley-locked band at $k_x = 0.67\pi/a$ [denoted by the blue dot in Fig. \ref{fig:fig2}(e)] and the pseudospin-locked band at $k_x = 0.05\pi/a$ [denoted by the orange dot in Fig. \ref{fig:fig2}(g)], respectively, showing that both the acoustic pressure and phase are uniformly distributed within the supercell, illustrating that these states are anomaly bulk states.
In Section 1 of the Supplementary Materials \cite{SM}, we further demonstrate that these anomaly bulk states (including valley, pseudospin-up, and pseudospin-down states) can be selectively excited for unidirectional propagation due to the chiral and helical properties.

\begin{figure}[h!]
\centering
\includegraphics[width=\columnwidth]{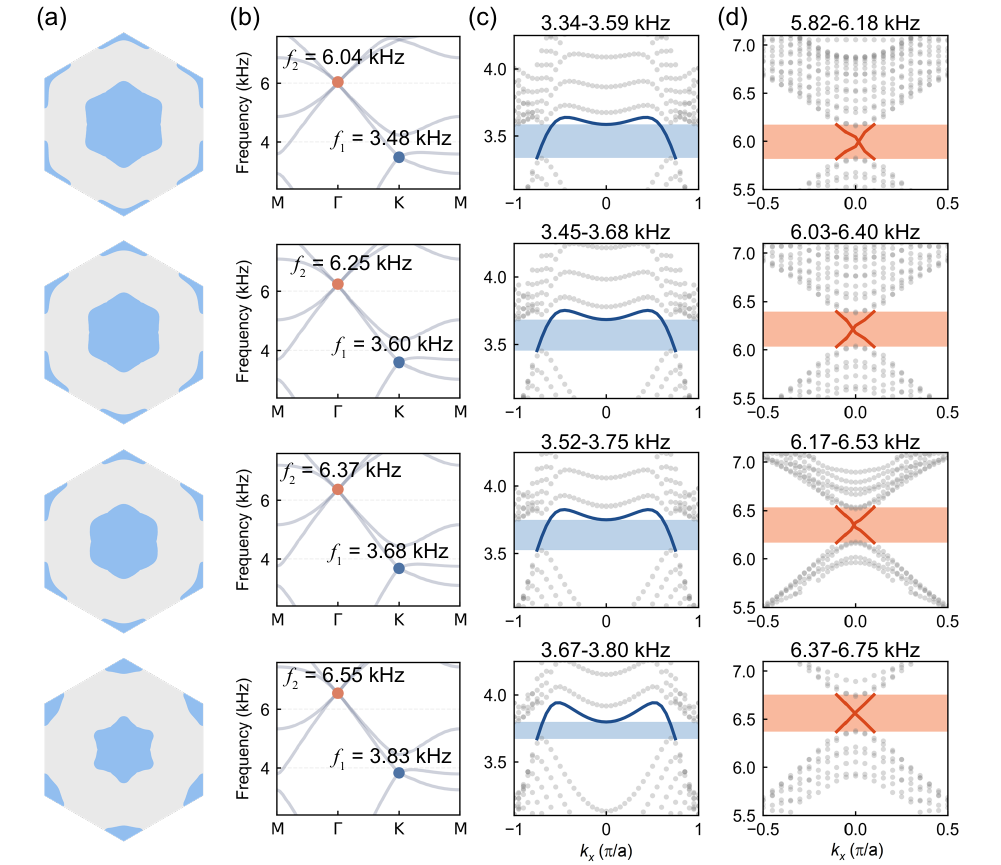} 
	\caption{(a) Schematic of the optimized UCs. (b) Band structures of the UCs. The blue (orange) dots indicate the degenerate points at which the valley (pseudospin) Hall gap opens. (c,d) Band dispersions of the supercells, showing the topological anomalous bulk states within the (c) valley Hall gap and (d) pseudospin Hall gap.}
	 \label{fig:fig-UCs}
\end{figure}

Meanwhile, our inverse-design method also allows the operating frequencies of the anomaly bulk states to be tuned flexibly, since these frequencies are directly set by the two target degeneracy frequencies $f_1$ and $f_2$ (corresponding to $\omega_1$ and $\omega_2$ in Appendix A) specified during optimization.
Figure \ref{fig:fig-UCs}(a) shows four UCs designed for different target frequencies.
As shown in Fig. \ref{fig:fig-UCs}(b), $f_1$ increases from 3.48 kHz to 3.80 kHz, while $f_2$ increases from 6.04 kHz to 6.55 kHz. The supercell dispersions in Figs.~\ref{fig:fig-UCs}(c) and \ref{fig:fig-UCs}(d) confirm that valley-locked and pseudospin-locked anomalous bulk states appear inside the corresponding gaps, and that their frequency windows shift upward with $f_1$ and $f_2$.
Apart from tuning the frequency of Dirac cones, the developed topology optimization method also enables the Dirac cone degenerated by higher-order bands, as discussed in Appendix C.
In addition, we discuss how the gap width evolves with the number of layers in Section 2 of the Supplementary Materials \cite{SM}.

To validate the valley- and pseudospin-locked anomaly bulk states, we construct a structure consisting of the periodically arrayed supercells [Fig. \ref{fig:fig2}(d)]
along the horizontal direction and fabricate it via the 3D printing, as shown in Fig. \ref{fig:fig3}(a), with the inset showing a magnified view of a UC (see Section 3 of the Supplementary Materials \cite{SM} for the experimental setup).
A point source is placed on the left side of the sample (red star).
Figures \ref{fig:fig3}(b) and \ref{fig:fig3}(e) present the simulated absolute acoustic pressure field at 3.68 kHz and 6.41 kHz,
corresponding to the valley- and pseudospin-locked anomaly bulk states, respectively,
showing that, after propagating several periods, the acoustic pressure fields are uniformly distributed within the structure for both the two kinds of anomaly bulk states.
Then, we experimentally extract the valley- and pseudospin-locked dispersion bands
by measuring the acoustic pressure along the horizontal purple dashed line in Fig. \ref{fig:fig3}(a)
and performing the Fourier transform, as shown in Figs. \ref{fig:fig3}(c) and \ref{fig:fig3}(f),
which agree well with the numerically calculated bands denoted by the white points.
Furthermore, the Fourier transform of the measured acoustic pressure within the dashed purple box [Fig. \ref{fig:fig3}(a)] shows that the propagating acoustic waves at 3.68 kHz and 6.41 kHz are locked at the $K$ [Fig. \ref{fig:fig3}(d)] and $\Gamma$ [Fig. \ref{fig:fig3}(g)] points, respectively, confirming that the valley- and pseudospin-locked anomaly bulk states have distinct momentum space distributions.

\begin{figure}[h!]
\centering
\includegraphics[width=\columnwidth]{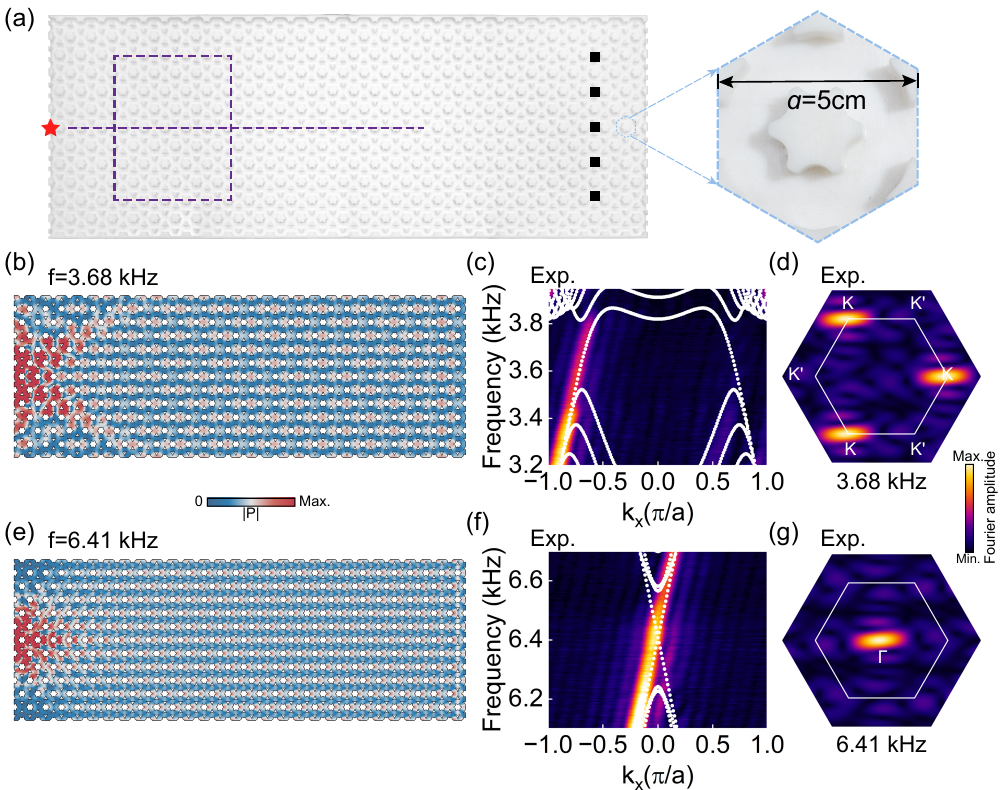} 
	\caption{(a) Sketch of the fabricated sample; the insert shows a magnified view of a UC.
	(b) Simulated absolute acoustic pressure field distribution at 3.68 kHz of the sample; the source is denoted by the red star in (a).
	(c) Experimentally measured dispersion by performing the Fourier transform on the measured field along the purple dashed line in (a) for the valley-locked anomaly bulk states.
	(d) Experimentally measured wave vector distribution at 3.68 kHz, obtained by conducting the Fourier transform on the measured field within the purple dashed box in (a).
	(e) Simulated absolute acoustic pressure field distribution at 6.41 kHz of the sample.
	(f) Experimentally measured dispersion by performing the Fourier transform on the measured field along the purple dashed line in (a) for the pseudospin-locked anomaly bulk states.
	(g) Experimentally measured wave vector distribution at 6.41 kHz, obtained by conducting the Fourier transform on the measured field within the purple dashed box in (a).}
	 \label{fig:fig3}
\end{figure}

\begin{figure*}[ht!]
\centering
\includegraphics[width=2\columnwidth]{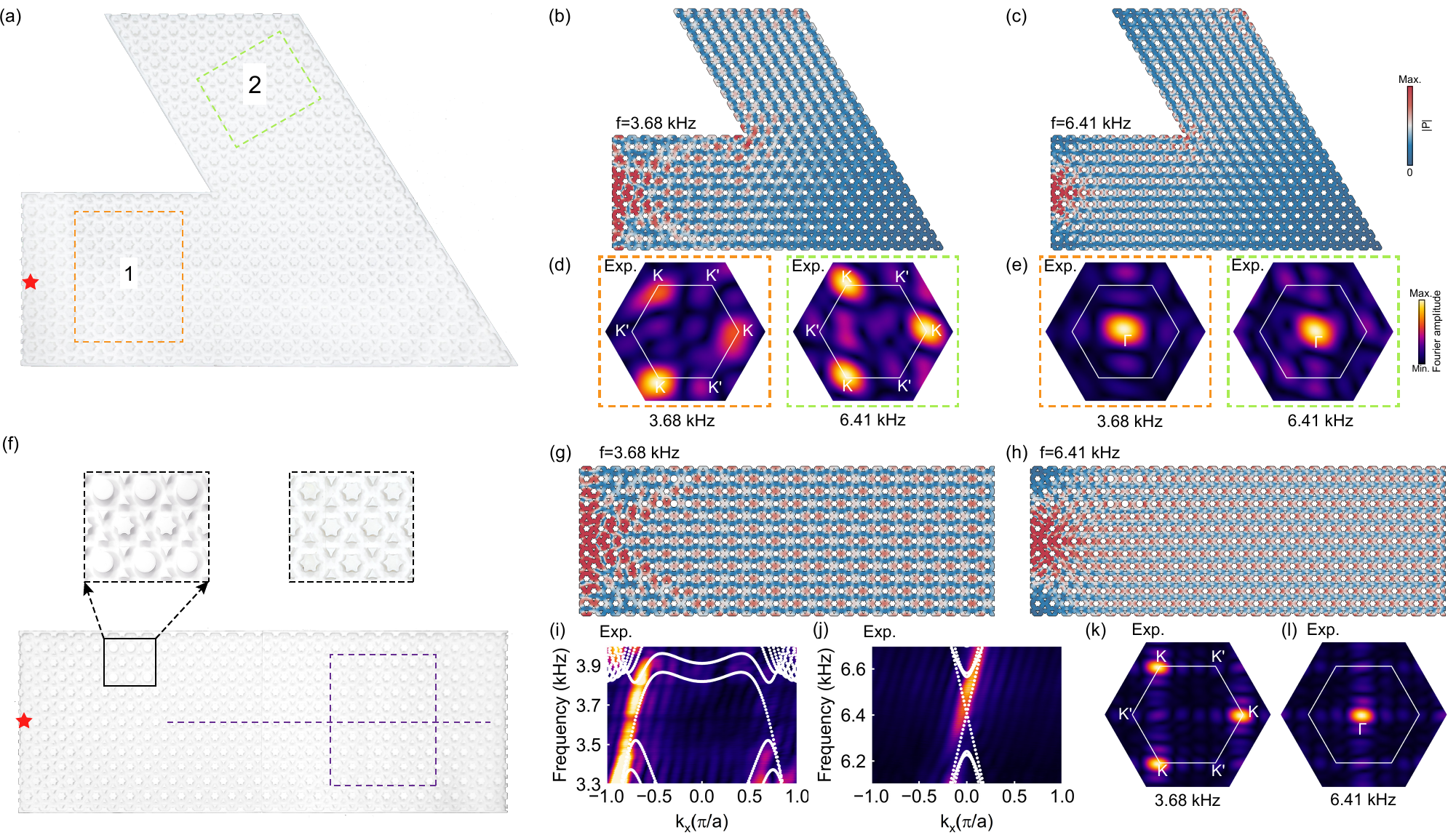} 
	\caption{(a) Photograph of the fabricated bend sample. (b, c) Simulated absolute acoustic pressure field at 3.68 kHz and 6.41 kHz in the bend sample. (d, e) The experimentally measured wave vector distribution at 3.68 kHz and 6.41 kHz in the dashed rectangle in (a); the left (right) panel corresponds to the orange (green) rectangle. (f) Photograph of the fabricated disordered sample. Inset picture: magnified views of a representative region in the pristine waveguide (right) and the disordered region (left). (g, h) Simulated absolute acoustic pressure fields at 3.68 kHz and 6.41 kHz in the disordered sample. (i, j) Experimentally measured dispersions (colors) of the disordered sample compared with the simulated bands (white dots) for the (i) valley- and (j) pseudospin- locked anomaly bulk states, respectively. (k,l) Experimentally measured wave vector distribution at 3.68 kHz and 6.41 kHz, obtained by conducting the Fourier transform on the measured field within the purple dashed box in (f).}
	 \label{fig:fig4}
\end{figure*}

With the momentum locked to the $K$ and $\Gamma$ points respectively, the valley- and pseudospin-locked anomaly bulk states are expected to exhibit robust transport, because backscattering requires coupling between states with distinct momentum-space characteristics.
To demonstrate such robustness, we build two structures containing a sharp bend [Fig. \ref{fig:fig4}(a)] and random disorders [Fig. \ref{fig:fig4}(f)], respectively.
Figures \ref{fig:fig4}(b) and \ref{fig:fig4}(c) show the simulated acoustic pressure fields at 3.68 kHz and 6.41 kHz within the bend structure, corresponding to the valley- and pseudospin-locked anomaly bulk states, respectively, demonstrating that, after passing the sharp bends, the acoustic pressure fields are still uniformly distributed without significant distortions.
Meanwhile, the measured momentum space distributions for the valley- and pseudospin-locked anomaly bulk states [Figs. \ref{fig:fig4}(d) and \ref{fig:fig4}(e)], extracted by performing the Fourier transform of the measured acoustic pressure within the dashed orange and green boxes, demonstrate that the momentum space distributions for these two types of anomaly bulk states remain undistorted before and after the waves propagate through the sharp bend; the valley- and pseudospin-locked states are locked at the $K$ and $\Gamma$ points, respectively.
The simulated fields within the structure with disorders at 3.68 kHz [Fig. \ref{fig:fig4}(g)] and 6.41 kHz [Fig. \ref{fig:fig4}(h)] also demonstrate that, after propagating several periods, the valley- and spin-locked acoustic waves are uniformly distributed without significant distortions. Meanwhile, the experimentally measured dispersion bands [Figs. \ref{fig:fig4}(i) and \ref{fig:fig4}(j)] and momentum space distributions [Figs. \ref{fig:fig4}(k) and \ref{fig:fig4}(l)] for the valley- and pseudospin-locked anomaly bulk states within the structure with disorders demonstrate that they are not distorted by the introduced disorders. Together, these results verify the robustness of the valley- and pseudospin-locked anomaly bulk states. 

\begin{figure}[h!]
\centering
\includegraphics[width=\columnwidth]{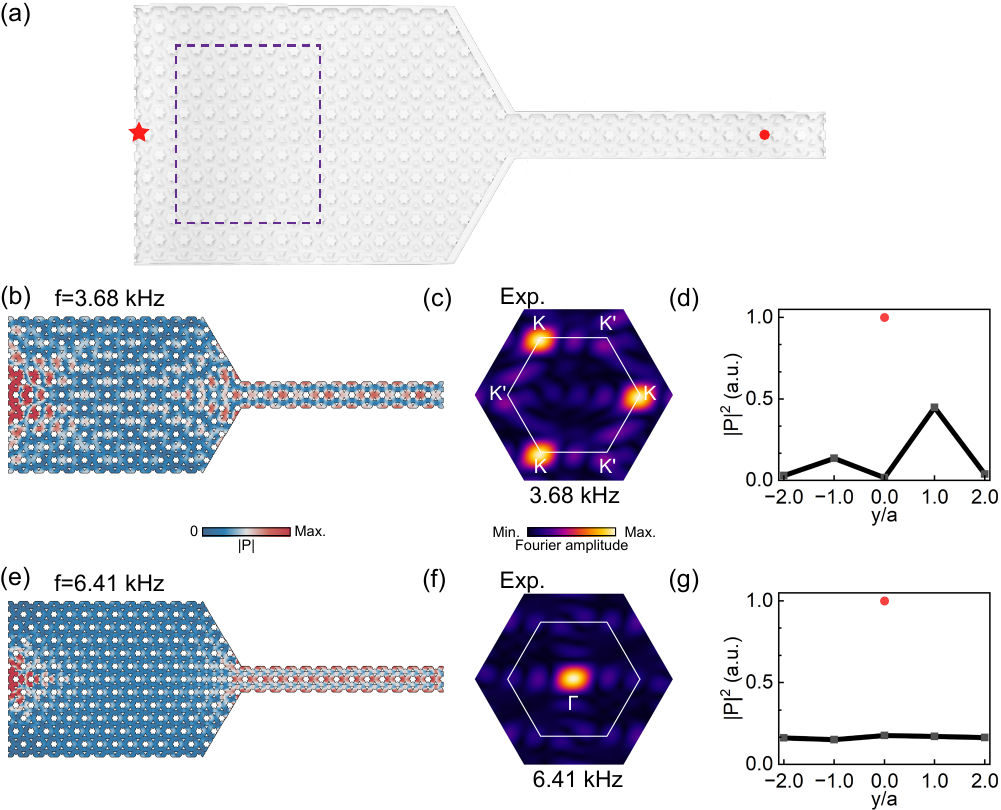} 
	\caption{(a) Photograph of the sample with a step change in the number of layers; the red star indicates the point source, and the purple dashed box delineates the region used for Fourier analysis. (b) Simulated absolute acoustic pressure field distribution at 3.68 kHz. (c) Experimentally measured wave vector distribution at 3.68 kHz, obtained by conducting the Fourier transform on the measured field within the purple dashed box in (a). (d) Experimentally measured intensities at 3.68 kHz at the locations marked by the black squares in Fig. \ref{fig:fig3}(a) and the red circle in Fig. \ref{fig:fig5}(a). (e) Simulated absolute acoustic pressure field distribution at 6.41 kHz. (f) Experimentally measured wave vector distribution at 6.41 kHz, obtained by conducting the Fourier transform on the measured field within the purple dashed box in (a). (g) Experimentally measured intensities at 6.41 kHz at the locations marked by the black squares in Fig. \ref{fig:fig3}(a) and the red circle in Fig. \ref{fig:fig5}(a).}
	 \label{fig:fig5}
\end{figure}

Apart from the robustness, the unique property of uniformly distributed phase fields [Figs. \ref{fig:fig2}(f) and \ref{fig:fig2}(h)] of the valley- and pseudospin-locked anomaly bulk states can be also exploited for focusing acoustic waves.
To demonstrate this property, we fabricate a sample with a step change in waveguide width, where the layer number is reduced from 12 to 2, as shown in Fig. \ref{fig:fig5}(a).
A point source is placed on the left (red star). From the simulated fields at 3.68 kHz and 6.41 kHz shown in Fig. \ref{fig:fig5}(b) and \ref{fig:fig5}(e), we can find that, both valley- and pseudospin-locked waves propagating along the wide channel are effectively squeezed into the narrow channel with enhanced acoustic pressure strength. Meanwhile, the momentum space distributions, extracted by performing the Fourier transform of the measured acoustic pressure within the purple dashed box in Fig. \ref{fig:fig5}(a), as shown in Fig. \ref{fig:fig5}(c) and 5(f), are not affected by the squeezing process, indicating that such process does not induce significant backscattering that distorts the momentum space distributions.
Figures \ref{fig:fig5}(d) and \ref{fig:fig5}(g) depict the measured acoustic field intensity at the point within the narrow channel [denoted by the red point in Fig. \ref{fig:fig5}(a)] with that at several reference points [denoted by the black squares in Fig. \ref{fig:fig3}(a)] within the straight sample with uniform waveguide width at 3.68 kHz and 6.41 kHz, respectively; the red point and black squares have the same horizontal distance to the left end of the respective sample.
We can find that, compared with the reference points, the measured pressure intensity within the narrow channel is significantly enhanced for both valley- and pseudospin-locked anomaly bulk states. 

\section{Conclusion and outlook} 
In conclusion, we achieve the integration of valley and pseudospin DOFs in one acoustic system. Through the developed topology optimization method, an acoustic metamaterial that simultaneously hosts a single Dirac cone at the $K$ point and a double Dirac cone at the $\Gamma$ point is inversely designed.
Via engineering its boundaries, both valley- and pseudospin-locked anomaly bulk states emerge within different frequency windows.
The dispersion spectrum and momentum space distribution of these anomaly bulk states are experimentally measured, for which the robustness is further demonstrated by introducing sharp bends and defects. Finally, the multifrequency energy enhancement effects for valley- and pseudospin-locked acoustic waves traversing a step waveguide are numerically and experimentally demonstrated.
From a theoretical standpoint, our results provide a recipe for engineering multiple topological DOFs within the same system, which can be also extended to non-Hermitian acoustic systems \cite{33,34} via the introduction of gain and loss.
From a practical perspective, the developed metamaterial provides a platform for engineering multiplexed acoustic devices with enhanced acoustic information processing capacities by encoding independent data streams across distinct topological DOFs.

\section*{Acknowledgments} 
This work is supported by the National Key R\&D Program of China (Grants No. 2022YFA1404400 and No. 2022YFA1404403), the National Natural Science Foundation of China (No. 92263208, 12102134), the Research Grants Council of Hong Kong SAR (AoE/P-502/20) and the Fundamental Research Funds for the Central Universities.\\
~

The authors declare no conflict of interest.\\
~

\section*{Data Availability} 
The data that support the findings of this article are openly available \cite{DATA}, embargo periods may apply.

\section{Appendix}

In this appendix, we provide the details to support the text.
In Sec. A, we present the topology optimization method for designing acoustic crystals hosting both single and double Dirac cones.
In Sec. B, we establish a theoretical framework for the parity-selection mechanism, which is discussed via a tight-binding model.
In Sec. C, we demonstrate that the developed topology optimization method can also be used to design Dirac cones degenerated by higher-order bands.

\subsection{The topology optimization method for designing acoustic crystals hosting both single and double Dirac cones} 
For source-free propagation, the pressure field distribution in an acoustic crystal is described by \cite{R1}:
\begin{equation}
-\nabla \cdot \left( \frac{1}{\rho(\mathbf{r})}\nabla p(\mathbf{r},\mathbf{k}) \right)
+ \frac{1}{B(\mathbf{r})}\frac{\partial^2 p(\mathbf{r},\mathbf{k})}{\partial t^2}
= 0
\label{eq:pressure-field}
\end{equation}
where $B(\mathbf{r})$ and $\rho(\mathbf{r})$ denote the bulk modulus and mass density, respectively, both satisfying the periodic relations
$B(\mathbf{r}) = B(\mathbf{r}+\mathbf{R})$ and
$\rho(\mathbf{r}) = \rho(\mathbf{r}+\mathbf{R})$.
Here, $\mathbf{R}$ is the lattice translation vector and $\mathbf{r}$ is the position vector.
Following the Bloch--Floquet theory \cite{R2,R3},
$p(\mathbf{r},\mathbf{k})$ can be formulated as
$p_{\mathbf{k}}(\mathbf{r},\mathbf{k})e^{i(\omega t+\mathbf{k}\cdot\mathbf{r})}$,
where
$p_{\mathbf{k}}(\mathbf{r},\mathbf{k})
= p_{\mathbf{k}}(\mathbf{r}+\mathbf{R},\mathbf{k})$,
with $\omega$, $\mathbf{k}=(k_x,k_y)$, and $p_{\mathbf{k}}$
denoting the angular frequency, Bloch wave vector, and periodic function, respectively.
By applying the finite element discretization, Eq. (\ref{eq:pressure-field}) can be formulated as
\begin{equation}
	\left(\mathbf{K}(\mathbf{k})-\omega^2\mathbf{M}\right)\mathbf{P}=0
\label{eq:dispersion-relation}
\end{equation}
where $\mathbf{P}$, $\mathbf{K}$, and $\mathbf{M}$ represent the nodal pressure vector, global stiffness matrix and global mass matrix, respectively. By sweeping the wave vector along the high-symmetry path of the first irreducible Brillouin zone and solving Eq. (\ref{eq:dispersion-relation}), the dispersion relations of the acoustic crystals can be obtained.

Here, we aim to design the $C_{6v}$-symmetric UC featuring a single Dirac cone at the $K/K'$ points and a double Dirac cone at the $\Gamma$ point. At the $K/K'$ points, the frequencies of the two eigenmodes used to form the single Dirac cone are denoted as $\omega_{K1}$ and $\omega_{K2}$, respectively. Similarly, at the $\Gamma$ point, the frequencies of the dipolar and quadrupolar modes used to form the double Dirac cone are denoted as $\omega_{d1,d2}$ and $\omega_{p1,p2}$, respectively. To simultaneously achieve the single Dirac cone at $\omega_1$ at the $K/K'$ point and the double Dirac cone at $\omega_2$ at the $\Gamma$ point, the optimization objective is set to minimize the following equation:

\begin{equation}
\begin{split}
g ={}& (\omega_{K1}-\omega_1)^2 + (\omega_{K2}-\omega_1)^2 + (\omega_{d1}-\omega_2)^2 \\
     & + (\omega_{d2}-\omega_2)^2 + (\omega_{p1}-\omega_2)^2 + (\omega_{p2}-\omega_2)^2
\end{split}
\label{eq:objective}
\end{equation}

which is equal to maximizing $f=-g$. Then, the UC is discretised with finite elements with each element denoted by the design variable $x_e$. So, the sensitivity of the objective function about the design variable $x_e$ can be formulated as

\begin{equation}
\begin{aligned}
\frac{\partial f}{\partial x_e}
= -2\Bigg[
&(\omega_{K1}-\omega_1)
\frac{\partial \omega_{K1}}{\partial x_e}
+(\omega_{K2}-\omega_1)
\frac{\partial \omega_{K2}}{\partial x_e}
\\
&+(\omega_{d1}-\omega_2)
\frac{\partial \omega_{d1}}{\partial x_e}
+(\omega_{d2}-\omega_2)
\frac{\partial \omega_{d2}}{\partial x_e}
\\
&+(\omega_{p1}-\omega_2)
\frac{\partial \omega_{p1}}{\partial x_e}
+(\omega_{p2}-\omega_2)
\frac{\partial \omega_{p2}}{\partial x_e}
\Bigg].
\end{aligned}
\label{eq:objective-sensitivity-x_e}
\end{equation}

where $\frac{\partial\omega_o}{\partial x_e}$ (o = $K1$, $K2$, $d1$, $d2$, $p1$ and $p2$) can be derived by differentiating both sides of Eq. (\ref{eq:dispersion-relation}):

\begin{equation}
\frac{\partial\omega_o}{\partial x_e} = \frac{1}{2\omega_o}\,
\mathbf{P}^T \left( \frac{\partial\mathbf{K}}{\partial x_e}
- \omega^2 \frac{\partial\mathbf{M}}{\partial x_e} \right)\mathbf{P}
\label{eq:sensitivity}
\end{equation}

The detailed calculation of Eq. (\ref{eq:sensitivity}) can be referred to Ref.\cite{R4}. After calculating the sensitivity of the objective function with respect to all elements, we adopt the bi-directional evolutionary structural optimization (BESO) method to iteratively update the design variables and maximize the objective function \cite{R5}. Once the objective function is maximized so that it approaches zero, the single Dirac cone is formed at the target frequency $\omega_1$, while the double Dirac cone is formed at the target frequency $\omega_2$. 

\subsection{Tight-binding model of boundary-induced valley and pseudospin bulk states} 

\begin{figure}[h!]
\centering
\includegraphics[width=\columnwidth]{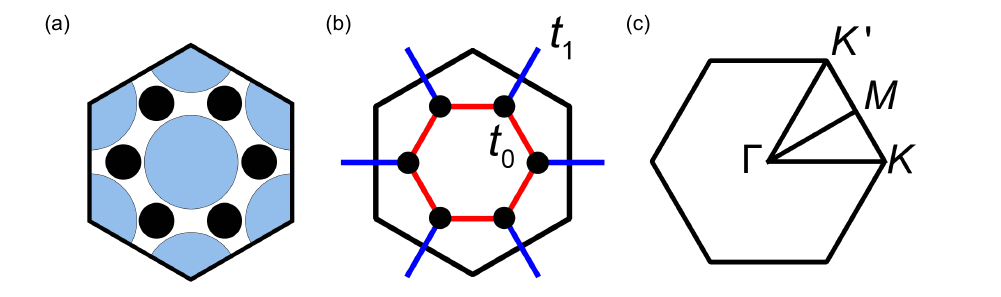} 
	\caption{\textit{}
	(a) Material distribution (blue) in the UC and the effective “particles” (black) describing the acoustic propagation. (b) The corresponding tight-binding model describing the acoustic wave propagation, where the intracell and intercell hoppings $t_0$ and $t_1$ can be tuned by modifying the dielectric distribution of the UC. (c) The first Brillouin zone and the corresponding high symmetry points.}
	 \label{fig:fig6}
\end{figure}

The physics of the boundary-induced hybrid topological waveguide supporting both quantum valley and pseudospin Hall states discussed in the main text could be well captured by the tight-binding model. The acoustic propagation due to the material distribution in the center and around the corners of the UC could be effectively described by six “particles” [see Fig. \ref{fig:fig6}(a)], whose intracell and intercell hoppings $t_0 $and $t_1$ [see Fig. \ref{fig:fig6}(b)] can be controlled by tuning the dielectric distribution around the center and corners of the UC, respectively. This model supports a double Dirac cone at the $\Gamma$ point of the Brillouin zone [see Fig. \ref{fig:fig6}(c)] when $t_0=t_1$ due to the mechanism of Brillouin zone folding \cite{R6}. Furthermore, the model also supports single Dirac cones at the $K/K'$ points of the Brillouin zone (see below), making this model suitable for the illustration of the physics leading to boundary-induced valley and pseudospin bulk states. 

We first discuss the boundary-induced quantum valley Hall bulk states of the model. The model supports Dirac cones at the $K/K'$ points of the Brillouin zone when $t_0=1$ and $t_1=2$, see Fig. \ref{fig:fig6}. The emergence of boundary-induced quantum valley Hall like bulk states around the Dirac point can be illustrated by a model in the continuum limit \cite{29WangCABS}.

\begin{figure}[h!]
\centering
\includegraphics[width=\columnwidth]{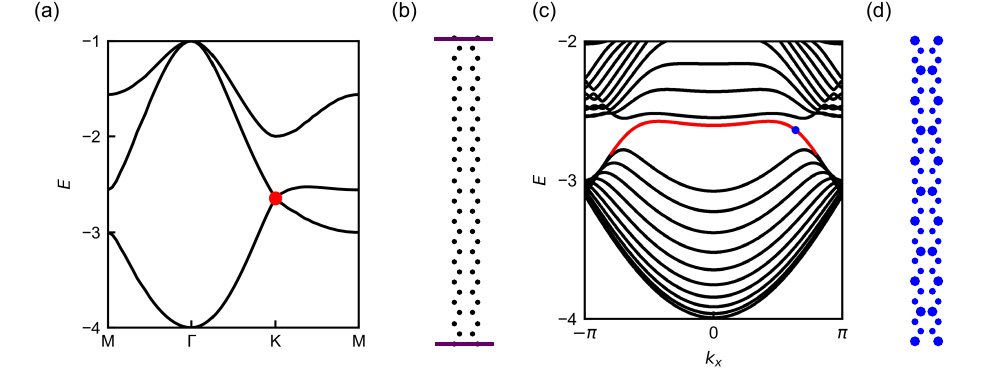} 
	\caption{
	(a) The tight-binding band diagram of the UC in Fig. \ref{fig:fig6} when $t_0=1$ and $t_1=2$, which hosts a Dirac cone at $K/K'$ (red dot). (b) The supercell for demonstrating boundary-induced quantum valley Hall bulk states, where the on-site potential at the top and bottom edges to emulate the acoustic hard boundary conditions is set to be $u=-0.5$. (c) The projected band diagram corresponding to the supercell in (b), which supports a band (red) of quantum valley Hall like states. (d) The eigenmode profile corresponding to the state marked by the blue dot in (c), demonstrating that the quantum valley Hall like state is a bulk state.} 
	 \label{fig:fig7}
\end{figure}

Around the Dirac point, the Hamiltonian could be described by:

\begin{equation}
H(y)=k_x\sigma_x-i\partial_y\sigma_y+m(y)\sigma_z
\label{eq:Hamiltonian}
\end{equation}

with the mass profile given by

\begin{equation}
m(y)=
\begin{cases}
m_2, & y \ge L,\\
0,   & 0 \le y \le L,\\
m_1, & y < 0.
\end{cases}
\label{eq:mass_profile}
\end{equation}

To ensure the confined nature of the solutions, the spinor wave function should be modified by $e^{|m_1|y}$ for $y<0$ and $e^{-|m_2|(y-L)}$ for $y>L$. Thus for $y<0$, the eigen-equation at $k_x=0$,$\omega=0$ is reduced to (by replacing $\partial_y$ by $|m_1|$) $\begin{pmatrix}
m_1 & -|m_1| \\
|m_1| & -m_1
\end{pmatrix}
\begin{pmatrix}
\psi_1 \\
\psi_2
\end{pmatrix}
=0$.
Using the fact that $m_1=\mathrm{sgn(}m_1\mathrm{)|}m_1|$ and ${|m}_1|=\mathrm{sgn(}m_1\mathrm{)\ }m_1$, one can readily obtain the eigen-solutions as $\begin{pmatrix}
\psi_1\\
\psi_2
\end{pmatrix}
=
\begin{pmatrix}
1\\
\operatorname{sgn}(m_1)
\end{pmatrix}$.
Thus, the eigen-solutions of Eq. (\ref{eq:Hamiltonian}) under the mass profile of Eq. (\ref{eq:mass_profile}) at $k_x=0$, $\omega=0$ can be summarized as

\begin{equation}
\psi=
\begin{cases}
e^{-|m_2|(y-L)}
\begin{pmatrix}
1\\
-\operatorname{sgn}(m_2)
\end{pmatrix}, & y>L,\\[6pt]
\begin{pmatrix}
1\\
\pm1
\end{pmatrix}, & L\ge y\ge 0,\\[6pt]
\psi'=e^{|m_1|y}
\begin{pmatrix}
1\\
\operatorname{sgn}(m_1)
\end{pmatrix}, & y<0.
\end{cases}
\label{eq:eigen_solutions}
\end{equation}

One can readily see that for the $\psi=\left(\begin{matrix}1\\\mathrm{1}\\\end{matrix}\right)$ solution within $L\geq y\geq0$, it requires that $m_1>0$ and $m_2<0$, whereas for the $\psi=\left(\begin{matrix}1\\\mathrm{-1}\\\end{matrix}\right)$ solution, one needs $m_1<0$ and $m_2>0$. This mechanism can be illustrated using the tight-binding model. To demonstrate this, we consider a supercell shown in Fig. \ref{fig:fig7}(b) and to emulate the acoustic hard boundary conditions, the on-site potential at the top and bottom edges of the supercell is set to be $u=-0.5$. The projected band diagram corresponding to this supercell is shown in Fig. \ref{fig:fig7}(c), from which one can see the emergence of a band of quantum valley Hall like states. Note this projected band diagram from the tight-binding calculations resembles closely with the one from numerical simulations of the acoustic materials as shown in Fig. \ref{fig:fig2}(e) of the main text. The representative eigenmode distribution of the emergent quantum valley Hall like state is shown in Fig. \ref{fig:fig7}(d), where one can see that the wave function distributes uniformly within the supercell, demonstrating it is indeed a bulk waveguide state.

\begin{figure}[h!]
\centering
\includegraphics[width=\columnwidth]{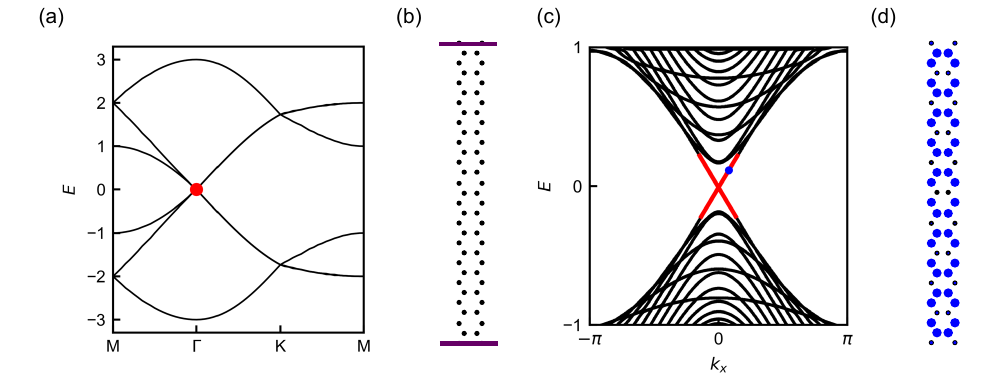} 
	\caption{
	(a) The tight-binding band diagram of the UC in Fig. \ref{fig:fig6}(b) when $t_0=1$ and $t_1=1$, which hosts a double Dirac cone at the $\Gamma$ point (red dot).
	(b) The supercell for demonstrating boundary-induced quantum pseudospin Hall bulk states, where the on-site potential at the top and bottom edges to emulate the acoustic hard boundary conditions is set to be $u=10$. (c) The projected band diagram corresponding to the supercell in (b), which supports two branches (red) of quantum pseudospin Hall like states. (d) The eigenmode profile corresponding to the state marked by the blue dot in (c), demonstrating that the quantum pseudospin Hall like state is a bulk state.} 
	 \label{fig:fig8}
\end{figure}

We now discuss the boundary-induced quantum pseudospin Hall bulk states of the model. The tight-binding model corresponding to the UC in Fig. \ref{fig:fig6}(b) supports a double Dirac cone at the $\Gamma$ point when the intracell and intercell hoppings are equal (see Fig. \ref{fig:fig8}(a)). To illustrate how the boundary conditions can select the quantum pseudospin Hall like states around the double Dirac cone, we begin with the following Hamiltonian \cite{R6}:

\begin{equation}
H(y)=
\begin{pmatrix}
H_{+}(y) & 0\\
0 & H_{-}(y)
\end{pmatrix}
\label{eq:Hamiltonian_pseudospin}
\end{equation}

with $H_{+}(y)=k_x\sigma_x-i\partial_y\sigma_y+m(y)\sigma_z$ and $H_{-}(y)=k_x\sigma_x+i\partial_y\sigma_y+m(y)\sigma_z$.
Using the same mass profile as in Eq. (\ref{eq:mass_profile}), and following similar derivations as above, we can obtain the eigen-solutions at
$k_x=0$, $\omega=0$ of Eq. (\ref{eq:Hamiltonian_pseudospin}), which are given by

\begin{equation}
\psi=
\begin{cases}
e^{-|m_2|(y-L)}
\begin{pmatrix}
1\\
-\operatorname{sgn}(m_2)\\
1\\
\operatorname{sgn}(m_2)
\end{pmatrix},
& y>L,\\[12pt]

\begin{pmatrix}
1\\
\pm 1\\
1\\
\pm 1
\end{pmatrix},
& L\ge y\ge 0,\\[12pt]

e^{|m_1|y}
\begin{pmatrix}
1\\
\operatorname{sgn}(m_1)\\
1\\
-\operatorname{sgn}(m_1)
\end{pmatrix},
& 0>y.
\end{cases}
\label{eq:eigen_solutions_pseudospin}
\end{equation}

To ensure the continuity of the solutions across $y=0$ and $y=L$, for $m_2>0$, one obtains $\psi=(1,-1,1,1)^T$ and $m_1<0$.
On the other hand, for $m_2<0$, one obtains $\psi=(1,1,1,-1)^T$ and $m_1>0$.
 So one can see, no matter whichever is the case, the spinor components of both the up and down sectors are ${(1,1)}^T$ and ${(1,-1)}^T$, which are opposite to each other, i.e., effectively forming the pseudospin up and down states. To demonstrate this mechanism in the tight-binding model, we consider a supercell shown in Fig. \ref{fig:fig8}(b) and to emulate the acoustic hard boundary conditions for the pseudospin physics, the on-site potential at the top and bottom edges of the supercell is set to be $u=10$. The projected band diagram corresponding to this supercell around the double Dirac cone is shown in Fig. \ref{fig:fig8}(c), from which one can see the emergence of two branches of quantum pseudospin Hall like states. Note that this projected band diagram from the tight-binding calculations also resembles closely with the one from numerical simulations of the acoustic materials as shown in Fig. \ref{fig:fig2}(g) of the main text. The representative eigenmode distribution of the emergent quantum pseudospin Hall like state is shown in Fig. \ref{fig:fig8}(d), where one can see that the wave function distributes uniformly within the supercell, demonstrating it is indeed a bulk waveguide state.

\subsection{Designing acoustic crystals with Dirac cone degenerated by higher-order bands}

\begin{figure}[h!]
\centering
\includegraphics[width=\columnwidth]{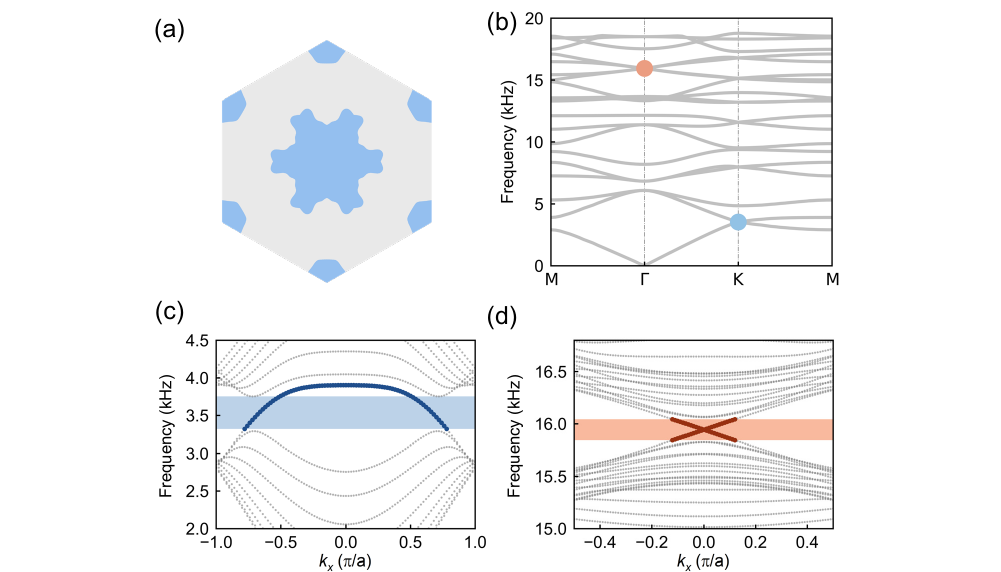} 
	\caption{
	(a) Schematic of the optimized UC. (b) Band structure of the UC. The blue (orange) dot indicates the single (double) Dirac point. (c,d) Band dispersions of the supercells, showing the (c) valley-locked and (d) pseudospin-locked anomaly bulk states.}
	 \label{fig:fig-HOUC}
\end{figure}

Apart from tuning the frequency of Dirac cones, the developed topology optimization method also enables the Dirac cone degenerated by higher-order bands. Figure \ref{fig:fig-HOUC} shows the topology-optimized UC [Fig. \ref{fig:fig-HOUC}(a)] with a single Dirac cone degenerated by the first and second bands [denoted by the blue point in Fig. \ref{fig:fig-HOUC}(b)] and a double Dirac cone degenerated by the fourteenth-seventeenth bands [denoted by the orange point in Fig. \ref{fig:fig-HOUC}(b)]. Figures \ref{fig:fig-HOUC}(c) and \ref{fig:fig-HOUC}(d) show the emergence of valley and pseudospin Hall anomaly bulk states based on the optimized UC, for which the frequency windows are 3.32-3.75 kHz and 15.8-16.0 kHz respectively. The latter is about five times that of the former, indicating that such a structure can be used for manipulating low- and high-frequency acoustic waves across widely separated frequency bands.

\FloatBarrier 
% \bibliography{ref}

\end{document}